\documentclass[aps,prd,twocolumn,superscriptaddress,showpacs,amsmath,amssymb,nofootinbib, preprintnumbers]{revtex4-1}
\usepackage{graphicx}
\usepackage{epstopdf}
\usepackage{float}
\usepackage{color}
\usepackage[toc,page]{appendix}
\usepackage[usenames,dvipsnames]{xcolor}
\usepackage[normalem]{ulem}

\newcommand{\be}{\begin{equation}}
\newcommand{\ee}{\end{equation}}
\newcommand{\ba}{\begin{eqnarray}}
\newcommand{\ea}{\end{eqnarray}}

\newcommand{\beq}{\begin{equation}}
\newcommand{\eeq}{\end{equation}}
\newcommand{\beqa}{\begin{eqnarray}}
\newcommand{\eeqa}{\end{eqnarray}}

\begin{document}
\title{Thermodynamics and Phase Transitions of NUTty Dyons}

\author{Alvaro Ballon Bordo}
\email{aballonbordo@perimeterinstitute.ca}
\affiliation{Perimeter Institute, 31 Caroline Street North, Waterloo, ON, N2L 2Y5, Canada}
\affiliation{Department of Physics and Astronomy, University of Waterloo,
Waterloo, Ontario, Canada, N2L 3G1}

\author{Finnian Gray}
\email{fgray@perimeterinstitute.ca}
\affiliation{Perimeter Institute, 31 Caroline Street North, Waterloo, ON, N2L 2Y5, Canada}
\affiliation{Department of Physics and Astronomy, University of Waterloo,
Waterloo, Ontario, Canada, N2L 3G1}

\author{David Kubiz\v n\'ak}
\email{dkubiznak@perimeterinstitute.ca}
\affiliation{Perimeter Institute, 31 Caroline Street North, Waterloo, ON, N2L 2Y5, Canada}
\affiliation{Department of Physics and Astronomy, University of Waterloo,
Waterloo, Ontario, Canada, N2L 3G1}

\date{March 29, 2019}

\begin{abstract}
Picking up the threads on a recent proposal \cite{Kubiznak:2019yiu}, we show how to formulate consistent thermodynamics of Lorentzian Taub--NUT spacetimes in the presence of electric and magnetic charges.
Namely, with an entropy identified with a quarter of the horizon area and no Misner time periodicity condition imposed, we show that a new pair of conjugate quantities can be introduced so that the NUT parameter 
can be independently varied and
the corresponding first law and Smarr relation can be consistently formulated. Moreover, we show that (contrary to the statements in the literature) the electric and magnetic parameters need not be proportional to one another and a full cohomogeneity first law including variations of both charges can be written down, provided one charge is considered on the horizon and the other at infinity. The
corresponding phase transitions are also briefly discussed.

\end{abstract}

\maketitle

\section{Introduction}

Celebrating its 78th birthday, the Lorentzian Taub--NUT solution has been a source of many puzzles over the years. Considered first as an anisotropic cosmological model \cite{taub1951empty}, and rediscovered later as a candidate for a black hole solution \cite{newman1963empty}, it features a string like singularity on the axis and
the accompanying spacetime regions with closed timelike curves. To avoid these pathologies, Misner \cite{misner1963flatter} proposed to periodically identify the time coordinate, rendering the string singularity unobservable. However, such an identification not only changes the topology of the solution and creates the timelike curves everywhere, but also makes the spacetime geodesically incomplete \cite{misner1963flatter, hawking1973large,hajicek1971causality}.

For these reasons the interest in the Lorentzian Taub--NUT solution gradually faded and the subsequent studies predominantly concentrated on the Euclidean case. Here the Wick rotated Taub--NUT solution provides an example of an interesting (special) Riemannian manifold, e.g. \cite{Page:1979aj, Page:1979zv}, and the thermodynamics with the Misner periodicity condition imposed can be consistently formulated \cite{Hawking:1998jf, Hawking:1998ct, Chamblin:1998pz, Emparan:1999pm, Mann:1999pc, Mann:1999bt, Johnson:2014xza, Johnson:2014pwa, Garfinkle:2000ms}. Such Euclidean thermodynamics demonstrates, however, a number of peculiar features. Namely, the entropy is not simply given by the horizon area over four but also picks up non-trivial contributions coming from Misner strings \cite{Hawking:1998jf, Hawking:1998ct, Carlip:1999cy, Garfinkle:2000ms}, the thermodynamic volume is potentially negative \cite{Johnson:2014xza, Johnson:2014pwa}, and, given that the NUT parameter cannot be independently varied due to the Misner condition, the first law is of reduced cohomogeneity.

The revival of the Lorentzian Taub--NUT solution began a few years ago, when it was shown by Cl{\'e}ment et al. \cite{Clement:2015cxa, Clement:2015aka, Clement:2016mll} that the Misner string singularity is far less pathological than previously expected. Specifically, the solution without imposing the Misner time periodicity condition was shown to be geodesically complete, as geodesics can freely pass through both the Misner string singularity \cite{Clement:2015cxa, Clement:2015aka} and the `black hole horizon' \cite{miller1971taub}. At the same time,
despite the presence of regions with closed timelike curves, no causality violations occur for geodesic observers  \cite{Clement:2015cxa, Clement:2015aka, Clement:2016mll}, suggesting that the Lorentzian Taub--NUT spacetimes with Misner strings present may actually be physical after all.

Further evidence supporting this idea was presented very recently in \cite{Kubiznak:2019yiu}, where it was shown that the Lorentzian Taub--NUT solution with Misner strings present is subject to sensible and consistent thermodynamics. In fact two suggestions for such thermodynamics were proposed: one where the entropy of the system is identified with a quarter of the horizon area and another where it is given by the Noether charge \cite{Garfinkle:2000ms}. In both scenarios a new charge (together with its conjugate quantity) is introduced so that the NUT parameter can be treated as an independent parameter and the first law attains full cohomogeneity. Both scenarios give the same manifestly positive thermodynamic volume that  obeys the reverse isoperimetric inequality \cite{Cvetic:2010jb}.
Of course, the two scenarios differ in the entropy formulae.  Moreover the new charges and their conjugate quantities are different, and so are the corresponding Smarr relations, see \cite{Kubiznak:2019yiu} for details.

In this paper we want to pick up the threads on these developments and show that the recently proposed thermodynamics can readily be extended to the charged Lorentzian Taub--NUT solution of the  Einstein--Maxwell theory \cite{brill1964electromagnetic, plebanski1976rotating, AlonsoAlberca:2000cs} (see also \cite{Mann:2005mb, Awad:2005ff, Dehghani:2006dk} for the higher-dimensional generalizations).
In general, such a solution admits both electric and magnetic charges, and we shall refer to it as a `Nutty dyon'.
The noteworthy feature of the presence of the NUT parameter is that it `mixes' the electric and magnetic parts of the electromagnetic field and, contrary to for example the dyonic Reisner--Nordstrom solution, the computation of the electric and magnetic charges depends on the radius of the two-sphere on which one integrates. As we shall see this interplay will play an important role for establishing the full cohomogeneity first law for these spacetimes.

{To formulate thermodynamics of these objects, we shall concentrate on the second scenario where the entropy is simply given by a quarter of the horizon area; that is, no entropy is associated to the Misner strings. This seems more plausible for several reasons:
i) As Misner strings are transparent for geodesics, there seems no opportunity for them to `hide any information' and store entropy. ii) For a `pure NUT system', which is characterized by a finite temperature, the entropy does not vanish, contrary to the first scenario, and is manifestly positive \cite{Kubiznak:2019yiu}. iii) The new conjugate quantities can be assigned geometrical meaning, e.g. \cite{Hawking:1998jf, Hawking:1998ct, Carlip:1999cy, Garfinkle:2000ms}, and are constrained by the action calculation \cite{Kubiznak:2019yiu}. }

Our paper is organized as follows. In the next section we introduce the Nutty dyon and discuss its basic properties.
{Sec.~III is devoted to the study of thermodynamics of this solution. Namely, in Sec.~\ref{Sec31} we explain the basic strategy for obtaining the thermodynamic quantities. Sec.~\ref{Sec32} presents thermodynamics upon imposing an extra `regularity condition' on the vector potential considered in the literature \cite{Awad:2005ff, Johnson:2014pwa, Dehghani:2006dk}. In Sec.~\ref{Sec33} we show that such a condition is perhaps not necessary and the full cohomogeneity first law can be obtained provided the magnetic charge is considered on the horizon and the electric one is calculated at infinity. Sec.~\ref{Sec34} presents the `magnetic version' of the derived thermodynamic laws, related to the `electric version' by the electromagnetic duality.
Sec.~\ref{Sec4} is devoted to the analysis of free energy and the discussion of possible phase transitions in the canonical (fixed charge) ensemble.
Sec.~\ref{Sec5} is devoted to conclusions. Appendix~\ref{AppA} recapitulates the thermodynamics of dyonic Reissner--Nordstrom solution, recovered in the limit of the vanishing NUT parameter.
}

\section{NUTty dyon}
In what follows we are going to consider the charged Lorentzian Taub--NUT solution of the Einstein--Maxwell theory with a cosmological constant
and both electric and magnetic charges switched on. The solution reads \cite{AlonsoAlberca:2000cs, Mann:2005mb, Awad:2005ff, Johnson:2014pwa}
\ba\label{j1}
ds^2&=&-f\bigl[dt+2n\cos\theta d\phi\bigr]^2+\frac{dr^2}{f}\nonumber\\
&&\qquad \quad +(r^2+n^2)(d\theta^2+\sin^2\!\theta d\phi^2)\,,\nonumber\\
A&=&-h(dt+2n\cos\theta d\phi)\,,
\ea
with
\ba\label{j2}
f&=&\frac{r^2-2mr-n^2+4n^2g^2+e^2}{r^2+n^2}-\frac{3n^4-6n^2r^2-r^4}{l^2(r^2+n^2)}\,,\nonumber\\
h&=&\frac{er}{r^2+n^2}+\frac{g(r^2-n^2)}{r^2+n^2}\,.
\ea
Here $n$ and $m$ are the NUT and mass parameters, $e$ and $g$ are the electric and magnetic parameters (see below), and $l$ stands for the AdS radius, $\Lambda=-3/l^2$.

The horizon is located at the largest root $f(r_+)=0$. This is a Killing horizon of the Killing vector
\be\label{xi}
\xi=\partial_t\,.
\ee

The parameters $e$ and $g$ are related to the electric and magnetic charges. Namely, employing the Gauss's law, the electric charge $q$ and the magnetic charge $q_m$ are given by
\be
q=\frac{1}{4\pi} \int_{S^2} *F\,,\quad q_m=\frac{1}{4\pi}\int_{S^2} F\,,
\ee
where $F=dA$ is the corresponding field strength.
Interestingly, performing the integration over a sphere of radius $r$, we find
\be\label{qqm}
q=\frac{e(r^2-n^2)-4gr n^2}{r^2+n^2}\,,\quad q_m=-\frac{2n\bigl(er+g[r^2-n^2]\bigr)}{r^2+n^2}\,,
\ee
which, contrary to common wisdom and known examples of dyonic solutions (Reissner--Nordstrom dyon for example) depends on the radius of the sphere. In particular, the asymptotic ($r\to \infty$) charges are
\be\label{QQm}
Q=e\,,\quad Q_m=-2gn\,.
\ee
However, as we shall see later these do not capture the whole story. {Note also the `electromagnetic duality', $q\leftrightarrow q_m$, by
replacing
\be\label{EM}
e\leftrightarrow -2ng\,,\quad 2ng\leftrightarrow e\,.
\ee
}

To recover the dyonic Reissner--Nordstrom solution, one has to carefully switch off the NUT parameter, while keeping the asymptotic charges $Q$ and $Q_m$ fixed, see appendix~\ref{AppA} for details.

Let us also note here the `regularity condition' imposed in the literature, see \cite{Awad:2005ff, Johnson:2014pwa, Dehghani:2006dk}, namely, that the vector potential $A$ has to vanish on the horizon. This requires $h(r_+)=0$ and yields
\be\label{gg}
g=-\frac{er_+}{r_+^2-n^2}\,.
\ee
That is, the magnetic parameter $g$ and the electric parameter $e$ are required to be proportional one another. Consequently the first law presented in \cite{Dehghani:2006dk, Johnson:2014pwa} is of reduced cohomogeneity as only one charge is independent. However, as we shall see in Sec. \ref{Sec33} this condition is not really needed for formulating consistent thermodynamics and in what follows \eqref{gg} will not be in general imposed.

Let us finally mention that similar to its uncharged cousin, the solution \eqref{j1} features Misner string singularities that are `symmetrically distributed' along the north pole and south pole polar axes. To avoid their presence, one could impose the Misner time periodicity condition,
\be\label{misner}
t\sim t+8\pi n\,.
\ee
However, as mentioned in the introduction this is problematic in many ways and we will not do so in this paper. That is, in what follows we keep the Misner strings present and consider $t$ non-compact.

\section{Thermodynamics}\label{Sec3}

\subsection{Strategy}\label{Sec31}
To formulate the thermodynamics of the Lorentzian NUTty dyon, we follow the strategy developed in \cite{Kubiznak:2019yiu}. We do not impose \eqref{misner} and seek a full cohomogeneity first law where all physical parameters $\{e,g, r_+, n, l\}$ can be independently varied. Such a first law should also be accompanied by the corresponding Smarr formula, with the two related by the dimensional scaling argument, e.g. \cite{Kastor:2009wy}.

To start with, we identify the temperature of the system with the temperature of the horizon,
\be\label{T}
T=\frac{f'}{4\pi}=\frac{1}{4\pi r_+}\Bigl(1+\frac{3(r_+^2+n^2)}{l^2}-\frac{e^2+4n^2g^2}{r_+^2+n^2}\Bigr)\,,
\ee
and assume the entropy to be given by the horizon area law:
\be\label{S}
S=\frac{\mbox{Area}}{4}=\pi (r_+^2+n^2)\,.
\ee

The thermodynamic mass can be calculated, for example, by using the conformal method \cite{Ashtekar:1999jx} and is simply given by
\be\label{m}
M=m\,,
\ee
while the total angular momentum of the spacetime vanishes.

The next step is to identify the thermodynamic electrostatic potential $\phi$. This is given by the difference of $-\xi\cdot A$ evaluated on the horizon minus the corresponding value at infinity:
\be
\phi=-\bigl(\xi\cdot A|_{r=r_+}-\xi\cdot A|_{r=\infty}\bigr)\,.
\ee
Employing \eqref{xi} for the generator of the horizon $\xi$ together with \eqref{j1} for the vector potential $A$, we find
\be\label{phi}
\phi=h(r_+)-g=\frac{er_+-2gn^2}{r_+^2+n^2}\,.
\ee
By employing the electromagnetic duality \eqref{EM}, we recover
\be\label{phim}
\phi_m=-\frac{n(2gr_++e)}{r_+^2+n^2}\
\ee
for the thermodynamic magnetostatic potential.

To proceed further we calculate the Euclidean action
\ba
I&=&\frac{1}{16\pi}\int_{M}d^{4}x\sqrt{g}\left[ R+\frac{6}{\ell^{2}}-F^2\right]
\nonumber\\
&&+ \frac{1}{8\pi}\int_{\partial M}d^{3}x\sqrt{h}\left[\mathcal{K}
-   \frac{2}{\ell} - \frac{\ell}{2}\mathcal{R}\left( h\right) \right]\,,
\label{action}
\ea
and identify it with the corresponding free energy  ${\cal G}=I/\beta$, with $\beta=1/T$ the periodicity of the Euclidean time coordinate.
Here, $\mathcal{K}$ and $\mathcal{R}\left( h\right)$ are respectively
the extrinsic curvature and Ricci scalar of the boundary, and we have included, apart from the Einstein--Hilbert--Maxwell--$\Lambda$ and Gibbons--Hawking terms, also the standard AdS counter-terms \cite{Emparan:1999pm}.
Note that in order to keep the metric and the vector potential real in the process of calculating the action one has to Wick rotate all of the following: the time coordinate, the NUT parameter, and the electric and magnetic charge parameters. The last three are at the end of the calculation Wick-rotated back, upon which we obtain the following result:
\ba\label{Free}
{\cal G}&=&\frac{m}{2}-\frac{r_+(3n^2+r_+^2)}{2l^2}-\frac{r_+e^2(r_+^2-n^2)}{2(n^2+r_+^2)^2}\nonumber\\
&&+\frac{2n^2r_+(r_+^2-n^2)g^2}{(r_+^2+n^2)^2}+\frac{4n^2 eg r_+^2}{(n^2+r_+^2)^2}\,.
\ea

The action as written corresponds to the grand-canonical (fixed electric potential and fixed asymptotic magnetic charge) ensemble \cite{Caldarelli:1999xj}.
According to \cite{Kubiznak:2019yiu}, this should now be identified with
\be\label{Fformula}
{\cal G}=M-TS-\phi Q - \psi N\,,
\ee
where $N$ is the new charge associated with the NUT parameter  and $\psi$ is its conjugate quantity \cite{Kubiznak:2019yiu}:
\be\label{psi}
\psi=\frac{1}{8\pi n}\,.
\ee
Thus, upon identifying $Q$, Eq. \eqref{Fformula} together with \eqref{Free} should yield the expression for the new charge $N$, which in its turn should satisfy the thermodynamic first law and the corresponding Smarr formula.

In what follows we proceed in two ways. First, we discuss `standard' thermodynamics where the extra condition \eqref{gg} is imposed and show that the corresponding (electrically charged) first law is straightforward to define. Second, we show that such a condition is in fact not necessary and a full cohomogeneity first law can be formulated, provided we consider one charge on the horizon and the other at infinity.

{Note finally that in the following we shall identify the cosmological constant $\Lambda$ with a thermodynamic pressure,
\be\label{P}
P=-\frac{\Lambda}{8\pi}=\frac{3}{8\pi l^2}\,,
\ee
and allow it to vary in the first law. The corresponding conjugate quantity
\be
V=\Bigl(\frac{\partial M}{\partial P}\Bigr)_{S,\dots}
\ee
is the thermodynamic volume (see \cite{Kubiznak:2016qmn} for more details) and for AdS black holes is conjectured to satisfy the reverse isoperimetric inequality \cite{Cvetic:2010jb}. 
}

\subsection{Electric first law}\label{Sec32}
Let us first revise the `standard thermodynamics' \cite{Awad:2005ff, Johnson:2014pwa, Dehghani:2006dk} where the condition  \eqref{gg}, $h(r_+)=0$, or
\be\label{gg2}
g=-\frac{er_+}{r_+^2-n^2}\,
\ee
is imposed.

Since the electric and magnetic parameters are no longer independent, {it may be possible to treat the solution as `only electrically charged', eliminating the dependence on $g$ via \eqref{gg2}.} The asymptotic electric charge and the electrostatic potential are then
\be
Q=e\,,\quad \phi=-g=\frac{er_+}{r_+^2-n^2}\,.
\ee
At the same time, the expression for the free energy \eqref{Free} significantly simplifies and yields
\be\label{FFF}
{\cal G}=\frac{m}{2}-\frac{r_+(3n^2+r_+^2)}{2l^2}-\frac{r_+e^2}{2(r_+^2-n^2)}\,,
\ee
while the temperature $T$, \eqref{T}, and the entropy $S$, \eqref{S}, are given by
\ba\label{TS2}
T&=&\frac{1}{4\pi r_+}\Bigl(1+\frac{3(r_+^2+n^2)}{l^2}-\frac{e^2(r_+^2+n^2)}{(r_+^2-n^2)^2}\Bigr)\,,\\
S&=&\pi (r_+^2+n^2)\,.
\ea
Eq.~\eqref{Fformula} with \eqref{psi} for $\psi$ then yields
\be
N=-\frac{4\pi n^3}{r_+}\Bigl(1+\frac{3(n^2-r_+^2)}{l^2}-\frac{e^2(3r_+^2+n^2)}{(n^2-r_+^2)^2}\Bigr)\,.
\ee

It is now easy to show that the standard first law and Smarr formula for charged solutions:
\ba
\delta M&=&T\delta S+\phi \delta Q+\psi dN+ V \delta P\,,\label{first1}\\
M&=&2(TS-VP+\psi N)+\phi Q\,, \label{Smarr1}
\ea
are satisfied, provided we set
\be\label{V}
V=\frac{4}{3}\pi r_+^3\Bigl(1+\frac{3n^2}{r_+^2}\Bigr)\,,
\ee
which is the same as for the uncharged case \cite{Kubiznak:2019yiu}. Thus we have recovered `standard thermodynamics' for electrically charged objects where all 4 parameters
$r_+, n, l$  and $e$ can be independently varied.

\subsection{Unconstrained thermodynamics}\label{Sec33}

No matter how beautiful the above result seems, there is something a bit disturbing about it. Namely, we have `only pretended' that the solution is only electrically charged. Note, however, that the asymptotic magnetic charge \eqref{QQm} does not vanish, and neither does the corresponding magnetostatic potential \eqref{phim}, they are given by
\be\label{Qmphim}
Q_m=\frac{2ner_+}{r_+^2-n^2}\,,\quad \phi_m=\frac{en}{r_+^2-n^2}\,.
\ee
So why do not these contribute in the first law \eqref{first1} and the Smarr relation \eqref{Smarr1}?

To answer this question, we should remember that the value of the magnetic charge $q_m$, \eqref{qqm}, depends on the radius of the sphere we integrate on: it is given by \eqref{Qmphim} at infinity and by
\be\label{Qmmm}
Q_m^{(+)}\equiv q_m(r=r_+)
\ee
on the horizon, which vanishes upon using the constraint \eqref{gg2}.
What if, it is this `horizon charge' $Q_m^{(+)}$ that should appear in the first law and the Smarr relation rather than the asymptotic charge $Q_m$ \eqref{Qmphim}? That is, is it possible that one could write
\ba
\delta M&=&T\delta S+\phi \delta Q+\phi_m \delta Q_m^{(+)}+\psi dN+ V \delta P\,,\label{first2}\\
M&=&2(TS-VP+\psi N)+\phi Q+\phi_m Q_m^{(+)}\,, \label{Smarr2}
\ea
instead of the relations \eqref{first1} and \eqref{Smarr1}?

To test this proposal, let us abandon the condition \eqref{gg2}---making the electric and magnetic parameters independent. In this case, $Q_m^{(+)}$ \eqref{Qmmm} becomes non-trivial, given by
\be\label{Qmplus}
Q_m^{(+)}=-\frac{2n(er_++g[r_+^2-n^2])}{r_+^2+n^2}\,,
\ee
and non-trivially contributes to the relations \eqref{first2} and \eqref{Smarr2}.
At the same time, the free energy is given by the full formula \eqref{Free} and so are the temperature $T$ \eqref{T}, and potentials $\phi$ \eqref{phi} and $\phi_m$ \eqref{phim}, while the entropy $S$ \eqref{S}, the mass $M=m$, and the charge $Q=e$ remain the same. The formula \eqref{Fformula} then implies the following expression for the charge $N$:
\ba
N&=&-\frac{4\pi n^3}{r_+}\Bigl(1+\frac{3(n^2-r_+^2)}{l^2}+\frac{(r_+^2-n^2)(e^2+4egr_+)}{(r_+^2+n^2)^2}\nonumber\\
&&\qquad\qquad-\frac{4n^2g^2(3r_+^2+n^2)}{(r_+^2+n^2)^2}\Bigr)\,.
\ea
It is now easy to verify that with these quantities the first law \eqref{first2} and the Smarr formula \eqref{Smarr2} are both satisfied provided we choose to be given by $V$ \eqref{V}.

We have thus found an unconstrained full cohomogeneity first law where all the physical parameters $\{e,g,r_+, l, n\}$ are independently varied. The slightly unconventional feature of this law is that the magnetic charge is considered on the horizon rather than at infinity. Note, however, that this does not really matter for more traditional spacetimes, a dyonic Reissner--Nordstrom solution for example, where the horizon and asymptotic magnetic charges simply coincide.

\subsection{Magnetic first laws}\label{Sec34}

The above result raises an interesting question: could we formulate an alternative full cohomogeneity first law, where the electric charge $q$ \eqref{qqm} would be taken on the horizon and the magnetic charge one at infinity? Namely, let us identify
\be\label{Qplus}
Q^{(+)}=q(r=r_+)=\frac{e(r_+^2-n^2)-4gr_+n^2}{r_+^2+n^2}\,,
\ee
and replace \eqref{Fformula} with
\be\label{Fformula2}
{\cal G}=M-TS-\phi Q^{(+)}-\psi \tilde N\,,
\ee
which yields
\be
\tilde N=N-16\pi n^2 \phi \phi_m\,.
\ee
Note that $N$ and $\tilde N$ are related by the electromagnetic duality \eqref{EM}.
It is then easy to verify that the following first law and Smarr relation hold:
\ba
\delta M&=&T\delta S+\phi \delta Q^{(+)}+\phi_m \delta Q_m+\psi d\tilde N+ V \delta P\,,\label{first3}\\
M&=&2(TS-VP+\psi \tilde N)+\phi Q^{(+)}+\phi_m Q_m\,, \label{Smarr3}
\ea
which is yet another unconstrained consistent formulation of the thermodynamics of NUTty dyons.

Note finally that, upon imposing $Q^{(+)}=0$, that is
\be\label{gg3}
e=\frac{4gr_+n^2}{r_+^2-n^2}\,,
\ee
the above formulas \eqref{first3} and \eqref{Smarr3} reduce to the `magnetic version' of the constrained thermodynamics:
\ba
\delta M&=&T\delta S+\phi_m \delta Q_m+\psi d\tilde N+ V \delta P\,,\label{first4}\\
M&=&2(TS-VP+\psi \tilde N)+\phi_m Q_m\,, \label{Smarr4}
\ea
related to the `electric version' studied in Sec.~\ref{Sec32} by the electromagnetic duality \eqref{EM}.
As in the electric version, no `horizon charges' appear in this formulation.

\begin{figure}
\begin{center}
\includegraphics[width=0.48\textwidth,height=0.3\textheight]{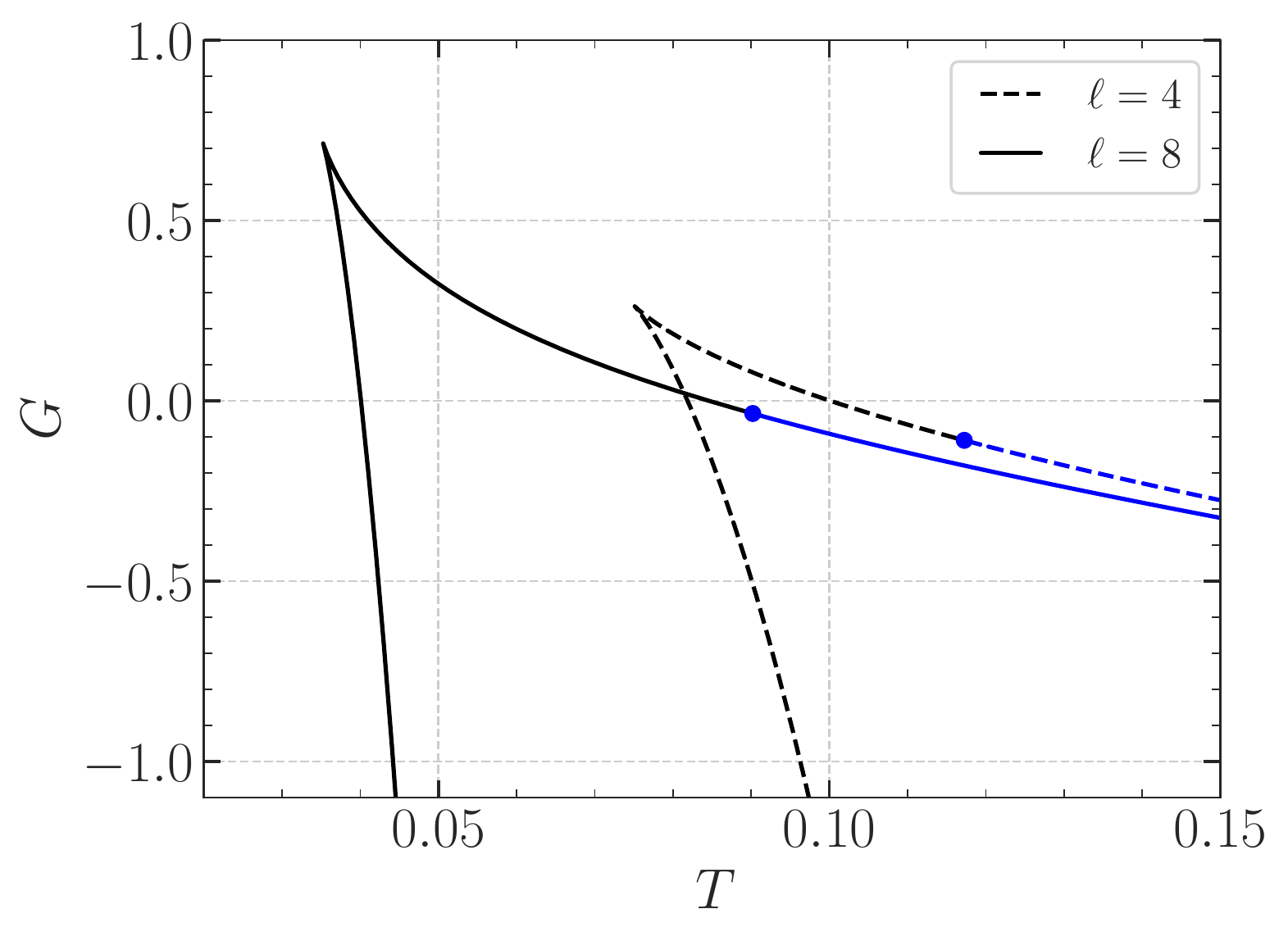}
\caption{{\bf Gibbs free energy: zero charge.} The $G-T$ diagram is displayed for fixed {$n=1$} and various pressures.
The presence of negative mass Taub--NUT solutions is indicated by blue curves that terminate at $M=0$ denoted by blue dots and smoothly join the corresponding positive mass solutions indicated by black curves. Although similar in form, the behaviour of the free energy no longer indicates the presence of a standard
Hawking--Page phase transition as no radiation phase with non-trivial $n$ exists.
}\label{fig:HP_e0}
\end{center}
\end{figure}

\section{Free energy and phase transitions}\label{Sec4}
\subsection{Choice of ensemble}

To understand the physics of the derived thermodynamic laws, let us look at the behavior of the free energy and uncover the possible phase transitions. For simplicity, we concentrate on the `purely electric case' studied in Sec.~\ref{Sec32} (with constraint \eqref{gg2} imposed).

It is well known \cite{Chamblin:1999tk, Kubiznak:2012wp} that in the absence of the NUT parameter charged AdS black holes demonstrate an interesting Van der Waals like behavior, when studied in the canonical (fixed charge) ensemble. Namely, the liquid/gas phase transition is mimicked by a small black hole to large black hole first order phase transition, with the corresponding coexistence line that emanates from zero pressure and temperature and terminates at a critical point characterized by the following critical temperature, black hole radius, and pressure \cite{Kubiznak:2012wp}:
\be
T_c^{(0)}=\frac{\sqrt{6}}{18\pi Q}\,,\quad r_c^{(0)}=\sqrt{6} Q\,,\quad P_c^{(0)}=\frac{1}{96 \pi  Q^2}\,.
\ee

Our aim is to  compare this behavior to the behavior of the charged Taub--NUT solution. Concentrating again on the canonical ensemble we have now, however, two possibilities: fixed $N$ ensemble and fixed $\psi$ ensemble. Since the former is algebraically more involved we concentrate here on the latter. The corresponding Gibbs free energy $G(T,P, \psi ,Q)$ is
\be
G=M-TS-\psi N\,,
\ee
and obeys
\be
\delta G=-S\delta T-N\delta \psi+V \delta P+\phi \delta Q\,.
\ee

In particular, we will study the behavior of this Gibbs free energy as a function of temperature, fixing the pressure, the charge, and the NUT parameter (and thence $\psi$). The global minimum then corresponds to the thermodynamically preferred phase.

\begin{figure}
\begin{center}
\includegraphics[width=0.48\textwidth,height=0.3\textheight]{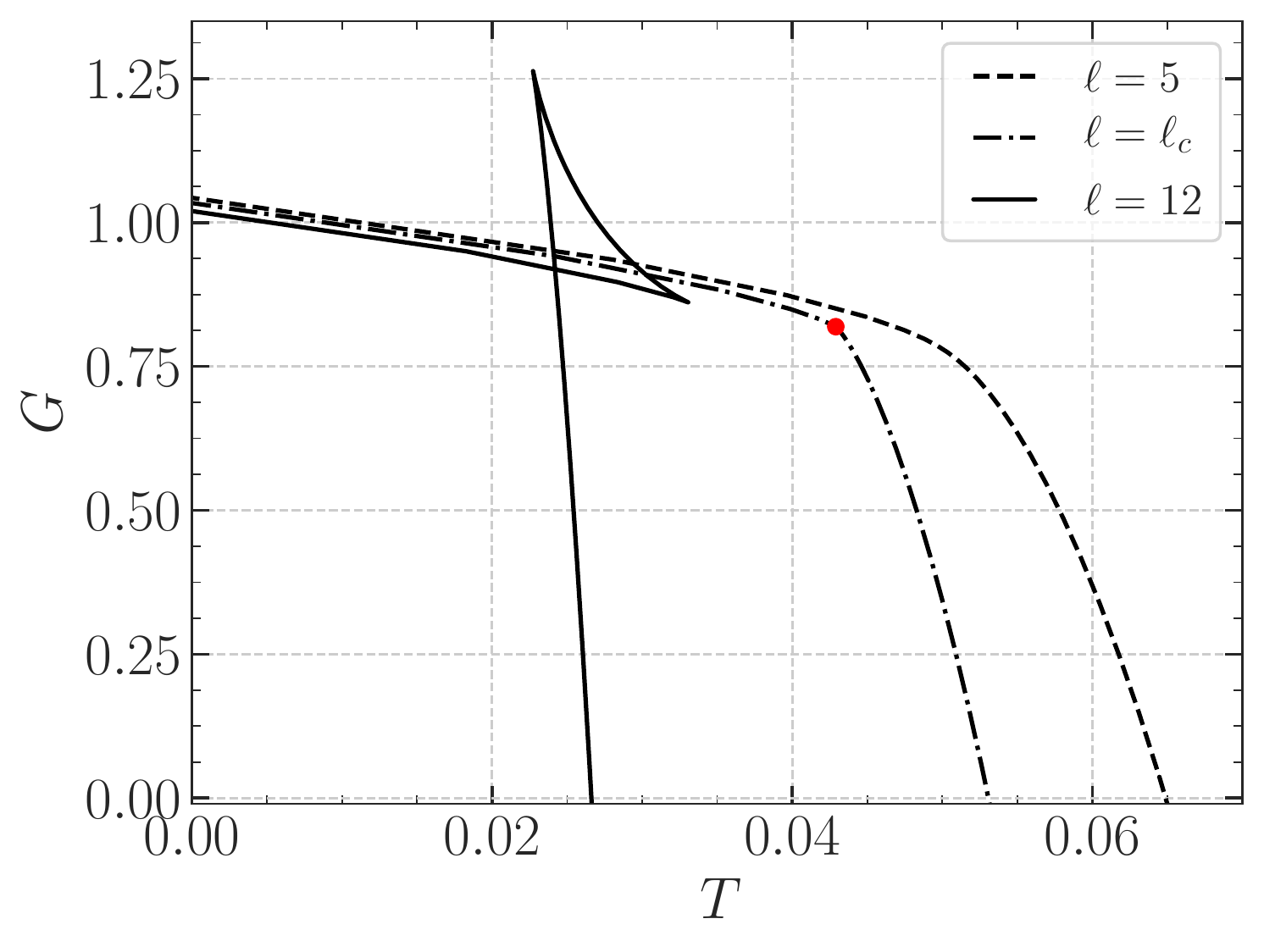}
\caption{{\bf Gibbs free energy: $P\approx P_c\ll P_t$ charged case.} The $G-T$ diagram is displayed for fixed $n=0.2$ and $Q=1$ and various pressures $P\ll P_t$, with $P_t$ given by \eqref{Pt}.  For pressures smaller than a critical value, $P<P_c$, given approximately by {Eq.~\eqref{eq:Pc}}, the Nutty dyon exhibits the classic swallowtail behavior indicating the small Taub--NUT to large Taub--NUT first order phase transition \`a la Van der Waals. The corresponding critical point is indicated by a red dot.
}\label{fig:SwallowTail}
\end{center}
\end{figure}

\subsection{Uncharged solution}

Concentrating first on the uncharged case $(Q=0$), we observe, see Fig.~\ref{fig:HP_e0}, a `Hawking--Page-like' behavior for the free energy  \cite{Hawking:1982dh}. However, there is an important difference. Since we work in a fixed $n\neq 0$ ensemble, there is no corresponding radiation phase and the similarity with the Hawking--Page is only superfluous (no phase transition occurs at the temperature where the free energy becomes negative).

{Moreover, it is a peculiar feature of the Taub--NUT solution that the horizon may exist also for a negative mass $M$. Such negative mass solutions are smoothly connected to the positive mass ones and there does not seem to be any strong objection against them describing a physical solution.\footnote{One objection against negative mass Taub--NUT solutions comes from the reverse isoperimetric inequality \cite{Cvetic:2010jb}. Namely, as shown in \cite{Kubiznak:2019yiu}, for $r_+<n/\sqrt{3+2\sqrt{3}}\approx 0.39 n$ the reverse isoperimetric inequality does not hold. If we want to preserve such an inequality, the mass cannot be too negative. }
Nevertheless, if one wants to restrict to positive mass solutions, a certain high temperature part of the upper branch has to be excluded, see blue curves in Fig.~\ref{fig:HP_e0}.}

\begin{figure}[H]

\includegraphics[width=0.48\textwidth,height=0.3\textheight]{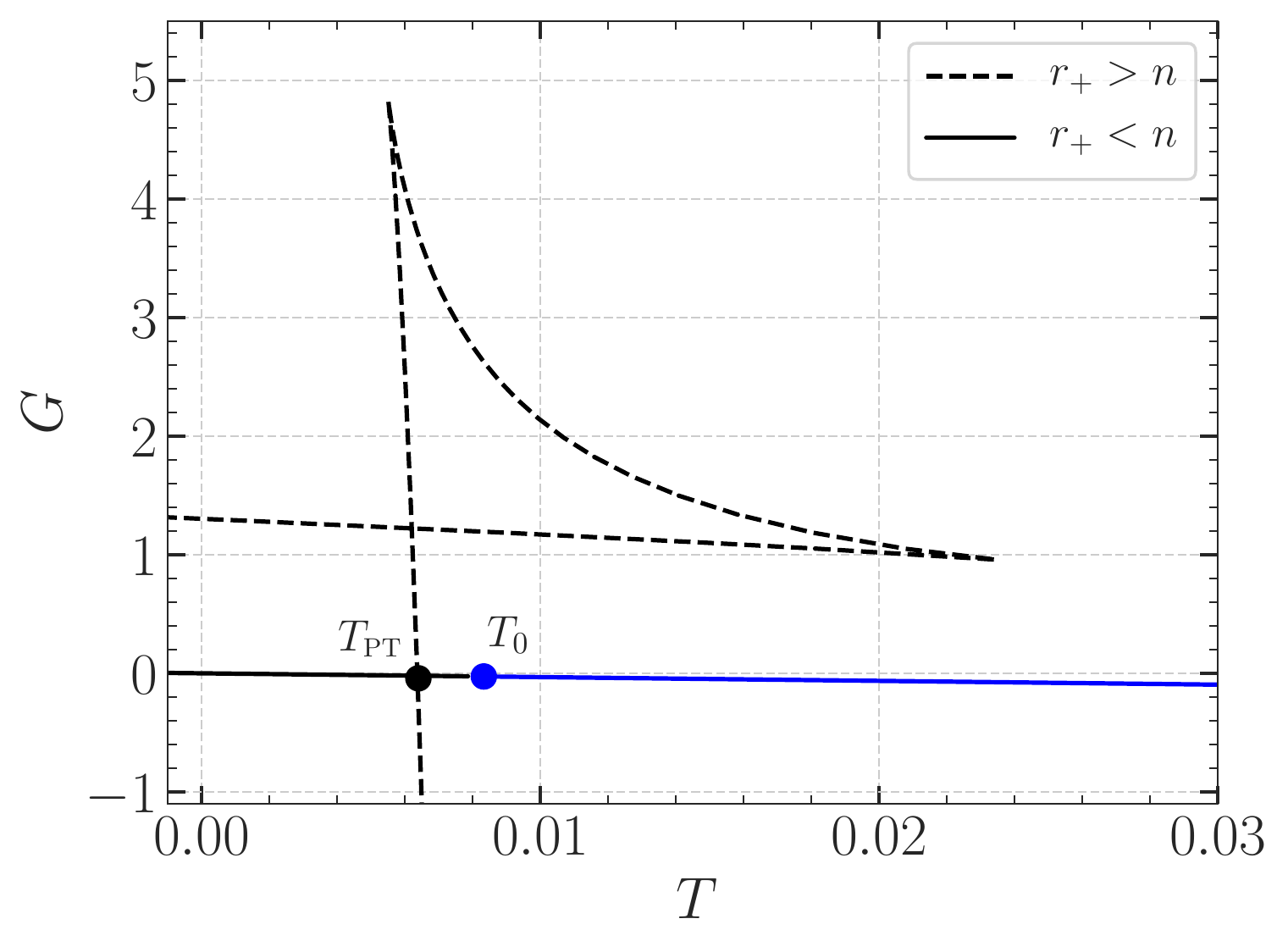}

\centering{(a)}
\includegraphics[width=0.48\textwidth,height=0.3\textheight]{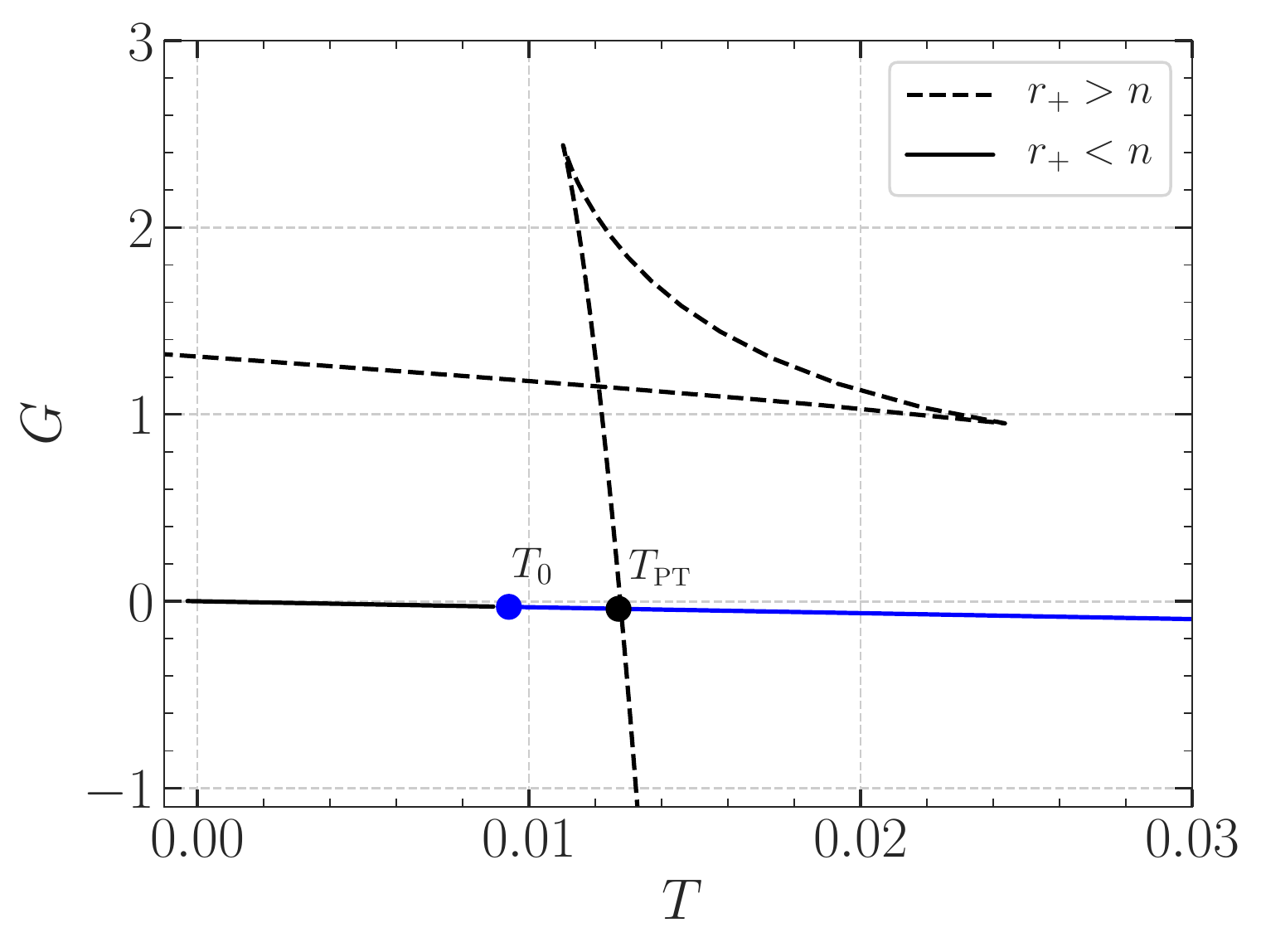}

\centering{(b)}
\caption{{{\bf Gibbs free energy: $P_t<P<P_c$ charged case.} The $G-T$ diagram is displayed for $n=1.006$, $Q=1$, and (a) $l=50$, (b) $l=25$.  Since $P>0>P_t$ we observe two Taub--NUT branches: the standard swallowtail branch of $r_+>n$  positive mass Taub--NUTs displayed by a dashed curve and the lower (solid curve) branch of $r_+<n$ possibly negative mass Taub--NUTs whose radius increases from right to left. As in the uncharged case the blue curve corresponds to negative mass solutions and terminates at a blue dot where $M=0$. As the pressure increases from (a) to (b), the relative position of the two important temperatures $T_0$ and $T_{\mbox{\tiny PT}}$ changes. If the negative mass solutions are disregarded, the first order phase transition at $T_{\mbox{\tiny PT}}$ no longer exists in (b). Instead, a zeroth order phase transition occurs at $T_0$ as the system jumps from small NUTs of the lower branch to intermediate NUTs of the swallowtail branch and is accompanied by the standard intermediate to large first order phase transition at a temperature corresponding to the bottom of the swallowtail.}}\label{fig:BHBranches}
\end{figure}

\begin{figure}
	\begin{center}
		\includegraphics[width=0.48\textwidth,height=0.3\textheight]{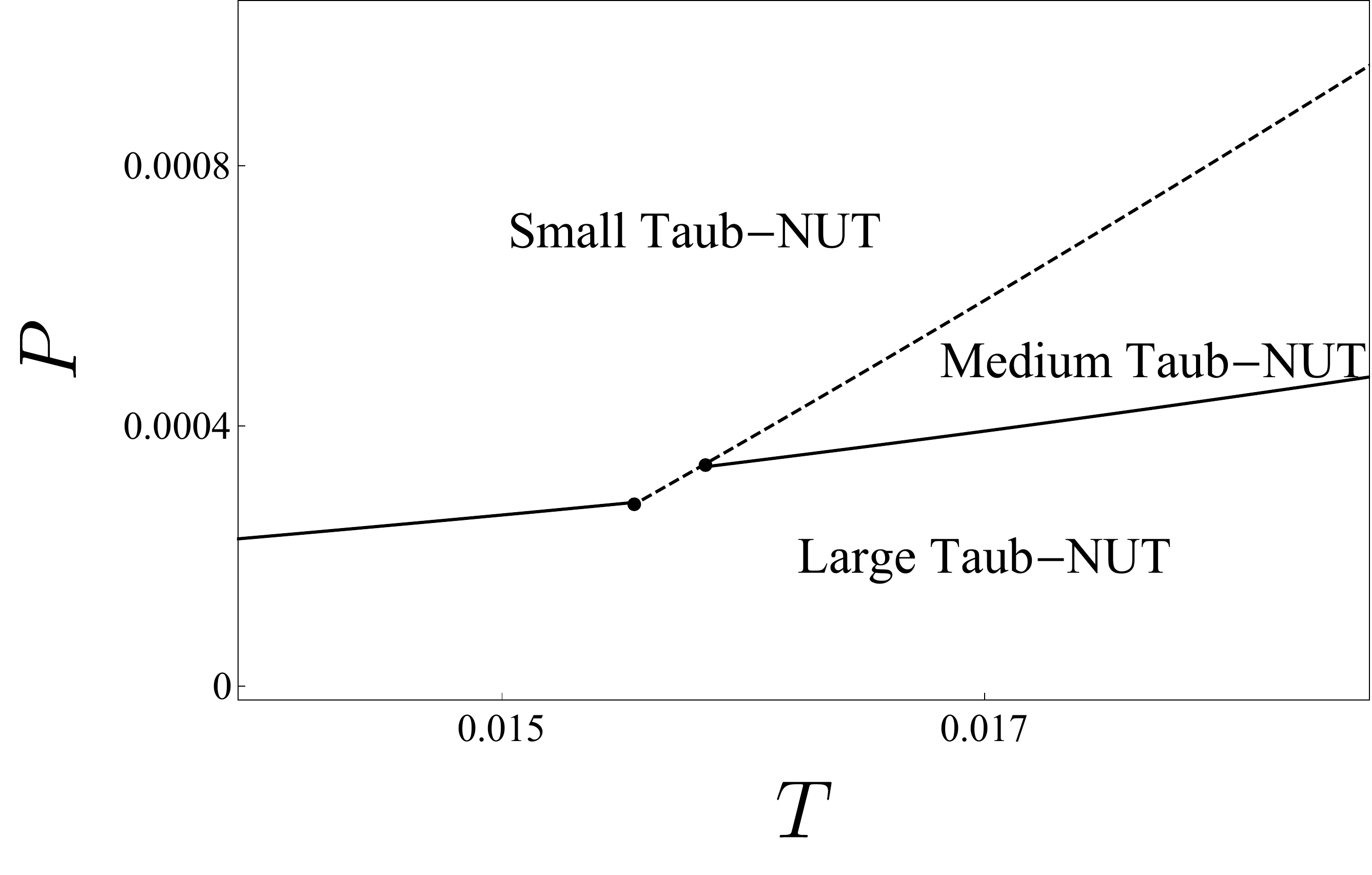}
		\caption{{{\bf Phase diagram.} The $P-T$ phase diagram is illustrated for $n=1$ and $Q=0.98$ and $P_t<P<P_c$ in the case when we disregard the negative mass solutions. The solid black lines indicate first order phase transitions whilst the dashed line indicates a zeroth order phase transition. The first order phase transition on the left corresponds to the intersection of the two branches at $T=T_{\mbox{\tiny PT}}<T_{0}$, Fig~\ref{fig:BHBranches}(a). For $P_t<0$ such a transition extends all the way to the point $(0,0)$. The zeroth order transition occurs when $T_{\mbox{\tiny PT}}>T_{0}$, Fig~\ref{fig:BHBranches}(b); increasing temperature causes a jump from the lower small Taub--NUT branch to the upper branch of the swallowtail. The first order phase transition on the right corresponds to the standard swallowtail phase transition and eventually terminates at a critical point. }}\label{fig:PT}
	\end{center}
\end{figure}

\subsection{Charged Taub--NUT}

The thermodynamic behavior of the charged Taub--NUT solution is a bit more complicated and interesting.

First of all, there is a branch of positive mass black holes that demonstrates qualitatively similar behavior to that of charged AdS black holes \cite{Kubiznak:2012wp}. Namely, there exists a critical pressure $P_c$ below which
the free energy displays a swallowtail behavior, indicating a small Taub--NUT to large Taub--NUT phase transition, \`a la Van der Waals, see Fig.~\ref{fig:SwallowTail}.

To find the corresponding critical point, we can use \eqref{P} and  rewrite the formula for the temperature \eqref{TS2}  into
the equation of state,
\be\label{eq:EoS}
P=\frac{ T}{2(r_+^2+n^2)}-\frac{1}{8\pi(r_+^2+n^2)}+\frac{Q^2}{8\pi(n^2-r_+^2)^2}\;.
\ee
Since we work in an ensemble where we fix $Q$ and the potential $\psi$ and thence also the NUT parameter $n$, $r_+=r_+(V)$ is given by inverting the volume equation \eqref{V}, and  the critical point occurs when
\be
\frac{\partial P}{\partial r_+}=0=\frac{\partial^2 P}{\partial r_+^2}\;.
\ee
Solving this perturbatively in $\tilde n =n/Q$ we find the following critical temperature, radius, and pressure:
\begin{align}
T_c&=T_c^{(0)}\Bigl(1-\frac{1}{4} \tilde n^2+\frac{83}{288}\tilde n^4+O(\tilde n^6)\Bigr)\;,\label{eq:Tc}\\
r_c&=r_c^{(0)}\Bigl(1+\frac{7}{12}\tilde n^2-\frac{515}{864} \tilde n^4+O(\tilde n^6)\Bigr)\;,\label{eq:rc}\\
P_c&=P_c^{(0)}\Bigl(1-\frac{2}{3}\tilde n^2+\frac{25}{27} \tilde n^4+O(\tilde n^6)\Bigr)\;.\label{eq:Pc}
\end{align}

However, this is not the whole story. Namely, for pressures that are greater than the following transition pressure:
\be\label{Pt}
P_t=\frac{Q^2-n^2}{8\pi n^4}\,,
\ee
there is yet another branch of small Taub--NUT solutions, whose radius increases as the temperature decreases, see Fig.~\ref{fig:BHBranches}. Such Taub--NUTs have positive mass
at small temperatures and more and more negative mass at high temperatures. We have verified that some Taub--NUTs of this branch violate the reverse isoperimeteric inequality~\cite{Cvetic:2010jb}. It remains to be seen whether demanding the validity of this inequality could remove the branch entirely.

Similar to the uncharged case, we can adopt two approaches: consider negative mass Taub--NUTs legitimate or consider them unphysical. Dependent on which approach we assume, we will observe different phase behavior.
In what follows we concentrate on
\be
P>P_t\,,
\ee
and briefly discuss the two options. Note that $P_t$ may be negative, in which case the new branch always exists.

In the first approach, there will always be a first order phase transition between large Taub--NUTs of the upper (swallowtail) branch and small Taub--NUTs of the lower branch; it  happens at a temperature
\be
T=T_{\mbox{\tiny PT}}\,,
\ee
where the two branches intersect, see Fig.~\ref{fig:BHBranches}. This is irrespective of whether the upper branch actually develops a swallowtail or not.

On the other hand, if only positive mass Taub--NUTs are taken seriously, only the low temperature (black curve) part of the lower branch is considered physical and the branch terminates at a finite temperature
\be
T=T_0\,,
\ee
where the mass of the Taub--NUT solution vanishes, denoted by blue dots in Fig.~\ref{fig:BHBranches}.
If such a temperature is bigger than $T_{\mbox{\tiny PT}}$, Fig.~\ref{fig:BHBranches}(a), the first order phase transition at $T_{\mbox{\tiny PT}}$ still takes place. However, if $T_0$ is smaller than $T_{\mbox{\tiny PT}}$, Fig.~\ref{fig:BHBranches}(b),  the corresponding first order phase transition at $T_{\mbox{\tiny PT}}$ no longer happens. Instead a zeroth order phase transition, see \cite{Altamirano:2013ane}, occurs at $T_0$ which may be accompanied by another (swallowtail origin) first order phase transition at higher temperatures. More specifically, increasing the temperature from below $T_0$, the system corresponds to a small Taub--NUT of the lower branch. At $T=T_0$, it jumps to the upper swallowtail branch of medium Taub--NUTs and follows its global minimum as the temperature is increased. In particular, if $T_0$ is smaller than the temperature of the bottom of the swallowtail (as is the case in Fig.~\ref{fig:BHBranches}(b)) there will be another, this time first order phase transition to a large Taub--NUT solution as $T$ is increased. The corresponding $P-T$ phase diagram is illustrated in Fig.~\ref{fig:PT}.

To summarize, the presence of the NUT charge changes qualitatively the thermodynamic behavior of charged solutions at pressures that are larger than $P_t$. Above this pressure a new branch of small Taub--NUTs appears and triggers a new type of behavior, especially if negative mass solutions are disregarded.

\section{Conclusions}\label{Sec5}

{
We have constructed consistent thermodynamics of the Lorentzian NUTty dyons---the dyonic Taub--NUT solutions of the Einstein--Maxwell theory, extending the previous results \cite{Kubiznak:2019yiu} to include electric and magnetic charges. The key ingredients in our construction were: i) to abandon the Misner time periodicity condition \eqref{misner}, keeping the Misner strings present, ii) to identify the entropy of the system with the quarter of the horizon area, and iii) to introduce a new conjugate pair of thermodynamic quantities $\psi-N$, upon which the full cohomogeneity first law can be restored.
}

{In the previous attempts at formulating the Euclidean thermodynamics of these solutions, \cite{Awad:2005ff, Johnson:2014pwa, Dehghani:2006dk}, in addition to the Misner condition an extra condition was imposed, requiring the vector potential to vanish on the horizon. This can be done in the Lorentzian case as well and leads to a constraint where the electric and magnetic parameters are not independent, one being proportional to the other, see \eqref{gg}. The corresponding `electric' thermodynamic laws are then easy to formulate, see Sec.~\ref{Sec32}. However, this is unsatisfactory for two reasons: i) the resultant first law is not of full cohomogeneity as only one electro/magnetic parameter can be varied independently and ii) only one of the electro/magnetic charges appears in the first law, despite that the Taub--NUT solution still carries both types of charges. }

To amend this situation, we have shown in Sec.~\ref{Sec33} that in fact the additional condition \eqref{gg} is not necessary and a full cohomogeneity first law, where all the parameters of the solution can be independently varied, can be formulated. The peculiar feature of this law is, however, that one of the charges is not an asymptotic charge and is instead evaluated on the horizon.
This might be related to the remarkable property of the NUTty dyon, namely that the NUT parameter mixes the electric and magnetic parts of the electromagnetic field and the calculation of the charges depends on the radius of the two sphere one integrates on. We note that in the limit of the vanishing NUT parameter, the horizon and asymptotic charges coincide and the discussed first law properly reduces to the standard one for the dyonic AdS black hole solution (see Appendix~\ref{AppA}).

To understand the physics of the derived thermodynamics, we have studied the free energy and uncovered the corresponding phase transitions. In particular, in order to compare with the Van der Waals behavior of the charged AdS black holes, we have concentrated on the canonical (fixed charge) ensemble with fixed potential $\psi$. We have shown that besides the standard swallowtail branch, there may exist another branch of small (possibly negative mass) Taub--NUTs which  significantly alter the thermodynamic behavior and the corresponding phase diagram; for example new first and zeroth order phase transitions may occur.  This, however, only happens for pressures that are higher than the transition pressure $P_t$, given by \eqref{Pt}. Below this pressure, the additional branch disappears and a behavior similar to that of the charged AdS black hole is observed.

To conclude, we have extended the Lorentzian thermodynamics of NUTs to the $U(1)$ charged case and explored their phase transitions, but we have not yet addressed the physical origin of the conjugate variables $\psi$ and $N$ related to the NUT parameter $n$. Moreover, it is an interesting open question whether the `mixed horizon/infinity first laws' are a specific feature innate to the NUTty dyon or whether similar laws can be formulated for other dyonic solutions. Finally, it remains to be seen whether these new phase transitions are unique to our method, or are present when one identifies the entropy with the Noether charge.

\section*{Acknowledgements}
The authors acknowledge the Perimeter Institute for Theoretical Physics.  Research at Perimeter Institute is supported by the Government of Canada through the Department of Innovation, Science and Economic Development Canada and by the Province of Ontario through the Ministry of Research, Innovation and Science.
D.K. is also grateful to the Natural Sciences and Engineering Research Council of Canada for their support.

\appendix

\section{The limit of vanishing NUT}\label{AppA}

In this appendix we will show how the well known thermodynamics of the dyonic AdS black holes is recovered in the limit of the vanishing NUT parameter.
In particular, we shall discuss the fate of the mixed horizon/infinity first laws \eqref{first2} and \eqref{first3} and what happens to the two conditions \eqref{gg2} and \eqref{gg3} in this limit.

To recover the dyonic AdS black hole solution we have to take the limit
\be
n\to 0\,,
\ee
while preserving both asymptotic electric and magnetic charges. This is achieved by holding
\be\label{fixed}
Q=e\,,\quad Q_m=-2ng=\hat g\,,
\ee
fixed, that is sending $g\to \infty$ in order to preserve $\hat g$.
The limit of the metric is straightforward and yields
\ba
ds^2&=&-fdt^2+\frac{dr^2}{f}+r^2(d\theta^2+\sin^2\!\theta d\phi^2)\,,\nonumber\\
f&=&1-\frac{2m}{r}+\frac{e^2+\hat g^2}{r^2}+\frac{r^2}{l^2}\,.
\ea
To get the right limit of the vector potential, we first add a pure gauge $A\to A+gdt$ and then take the limit $n\to 0$, to recover
\be
A=-\frac{e}{r}dt+\hat g \cos \theta d\phi\,,
\ee
which completes the dyonic AdS solution.
Note that the pure gauge term we added was used to regularize the divergence in this limit.
Note also that such a vector potential does not vanish on the horizon.

The free energy \eqref{Free} now becomes
\be
{\cal G} =\frac{m}{2}-\frac{r_+^3}{2l^2}-\frac{e^2-\hat g^2}{2r_+}\,,
\ee
which agrees with Eq. (33) in \cite{Caldarelli:1999xj}. Although the potential $\psi$ diverges, the products $\psi N$ as well as $\psi \tilde N$, which enter thermodynamic relations, both vanish. The mass is still given by $M=m$ and the pressure $P$ by \eqref{P}, while other thermodynamic quantities $T$ \eqref{T}, $S$ \eqref{S}, $\phi$ \eqref{phi}, $\phi_m$ \eqref{phim}, and $V$ \eqref{V}, now become
\ba
T&=&\frac{1}{4\pi r_+}\Bigl(1+\frac{3r_+^2}{l^2}-\frac{e^2+\hat g^2}{r_+^2}\Bigr)\,,\quad S=\pi r_+^2\,,\nonumber\\
\phi&=&\frac{e}{r_+}\,,\quad \phi_m=\frac{\hat g}{r_+}\,,\quad V=\frac{4}{3}\pi r_+^3\,.
\ea
Of course, in this case the horizon charges $Q^{(+)}$ \eqref{Qplus} and $Q_m^{(+)}$ \eqref{Qmplus} now become identical to the asymptotic charges $Q$ and $Q_m$:
\be
Q^{(+)}=Q=e\,,\quad Q_m^{(+)}=Q_m=\hat g\,.
\ee
This means that the formulae \eqref{Fformula} and \eqref{Fformula2} coincide to give
\be
{\cal G} =M-TS-\phi Q\,,
\ee
and the mixed horizon/infinity first laws \eqref{first2} and \eqref{first3} now become the standard first law for the AdS dyonic black hole solutions:
\be
\delta M=T\delta S+\phi \delta Q+\phi_m \delta Q_m+V\delta P\,.
\ee
This is accompanied by the corresponding Smarr relation
\begin{align}
M=2(TS-VP)+\phi Q+\phi_m Q_m\,,
\end{align}
which is a limit of either \eqref{Smarr2} or \eqref{Smarr3}. Thus we have shown that the unconstrained thermodynamics of the NUTty dyons properly reduces to the unconstrained thermodynamics of the dyonic AdS black hole.

Let us finally look at the constrained thermodynamics \eqref{first1} and \eqref{first4} and the related conditions \eqref{gg2} and \eqref{gg3}.
Starting from condition \eqref{gg2}, we see that its $n\to 0$ limit is inconsistent with keeping both asymptotic charges, \eqref{fixed}, fixed.
However, the electric charge $e$ can remain fixed, in which case the condition \eqref{gg2} yields $g$ finite and so $\hat g$ has to vanish. This simply corresponds to electrically charged AdS black holes. We have thus shown that the constrained `electric version' thermodynamics studied in Sec.~\ref{Sec32} yields the thermodynamics of electrically charged AdS black holes.

Similarly, starting from condition \eqref{gg3}, its $n\to 0$ limit is also inconsistent with keeping both asymptotic charges, \eqref{fixed}, fixed, but the magnetic charge $\hat g$ can remain fixed, in which case the condition \eqref{gg3} yields $e=0$. The `magnetic version' of NUTty thermodynamics thus reduces to the thermodynamics of magnetically charged AdS black holes. Of course, the two are related by the electromagnetic duality $e\leftrightarrow \hat g$.

\providecommand{\href}[2]{#2}\begingroup\raggedright\endgroup

\end{document}